\documentclass[11pt]{article}
\usepackage{geometry}
\usepackage[parfill]{parskip}
\usepackage{graphicx}

\date{\today \vskip 0.5cm  Contact: http://mics.uni.lu}
\title{Design and Implementation of a Master of Science in Information and Computer Sciences - An Inventory and retrospect for the last four years}
\author{Christoph Schommer, Director of Studies \\ Master of Science in Information and Computer Sciences (MICS)\\
Dept. of Computer Science and Communication\\ University of Luxembourg, Campus Kirchberg}

\makeindex

\begin{document}
\maketitle
\begin{abstract}
This Master of Science in Computer and Information Sciences (MICS) is an international accredited master program that has been initiated in 2004 and started in September 2005. MICS is a research-oriented academic study of 4 semesters and a continuation of the Bachelor towards the PhD. It is completely taught in English, supported by lecturers coming from more than ten different countries. This report compass a description of its underlying architecture, describes some implementation details and gives a presentation of diverse experiences and results. As the program has been designed and implemented right after the creation of the University, the significance of the program is moreover a self-discovery of the computer science department, which has finally led to the creation of the today's research institutes and research axes.
\end{abstract}

\newpage

\section{Beginning}\label{beginning}

When\footnote{All information written in the document bases on the information that is currently valid. Things like financial support or others may change. There is no warranty.} the University of Luxembourg has been created in the end of 2003 by a join of three existing but independent institutions, the study program of computer science did not exist but was hidden/integrated into an engineering program (\textsf{Ing\'enierie Industriel}) as a specialisation only. The study has been industry-oriented but less on science, generally being on an undergraduate level, highly oriented towards software design and programming but less on general fundamentals or research-related issues.

The self-approach of the University of Luxembourg has ever been research-oriented, and the aim of creating an university has ever been followed by the claim of establishing a research-oriented university. In this respect, the wish of designing and performing a master program in computer science has been born right after the creation of the University of Luxembourg.

Nevertheless, fundamental questions have come up, for example, are we able to perform the study program or can we meet the expectations, even though we have to face up to the fact that the university is young and inexperienced and that we not have that much students as other universities still possess? Continuously aware of this, we have identified major guidelines, namely

\begin{itemize}
    \item[{$\circ$}] to start from our existing strengths, i.e., to join forces to reach a critical mass.
    \item[{$\circ$}] to offer an interdisciplinary perspective.
    \item[{$\circ$}] to focus on scientifically promising niches.
    \item[{$\circ$}] to take into account the regional academic context.
    \item[{$\circ$}] to observe short- and long-term societal and economical needs.
\end{itemize}

and started the organisation of the program, the share of responsibilities, etc. The final implementation of \textsf{MICS} have been done it in the spirit of the Bologna agreement, accomplished with diverse marketing campaigns like a permanent presence on public events, staying in contact with alumni, and offering social activities to strengthen student life.

Today, we are sure to have prepared a master program that fits into the local industrial and academic landscape. The number of students is constantly around 15 per year, we still are in co-operation and collaboration with other universities, and share an intimacy with researchers from the departments, working in research projects and global research axes. The internal acceptance is still as high as four years ago; being a personal program of study, \textsf{MICS} is still the gateway to the doctorate.

\section{Architecture}\label{architecture}

\subsection{Positioning}\label{positioning}

\begin{figure}[htbp]
   \centering
   \includegraphics[width=8cm]{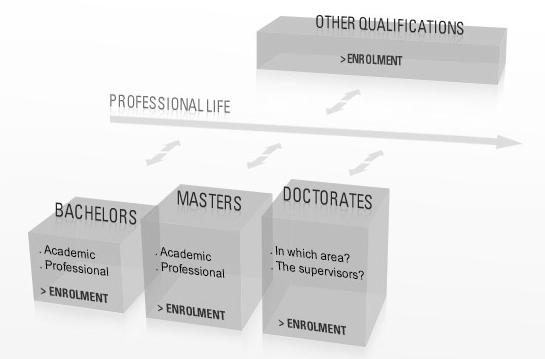} 
   \caption{From Bachelor to the Doctorate via the Master}
   \label{fig:archi0}
\end{figure}

The masters program is embedded in natural sequence of undergraduate studies (Bachelor), graduate studies (Master), and doctorate (Figure \ref{fig:archi0}). Both the Bachelor and the Master program can be either academic or professional/industrial. Following the conditions stated in the Bologna program (\cite{bologna}), the number of ECTS to may vary from 180 to 240 ECTS on Bachelor level and from 60 to 120 ECTS in the Master level.

For \textsf{MICS}, we have identified four semesters with 120 ECTS (sections \ref{courses}) in total. ECTS stands for \textsf{European Credit Transfer System} and refers to an European-wide education system that supports transparency, mobility and exchange between European universities. In \textsf{MICS}, each course has assigned a number of ECTS, where 1 ECTS corresponds to 15h global working time of a student per semester. The global working time compounds the course itself, the pre- and post-processing of a course, and the preparation for the examination. In general, a course in \textsf{MICS} may start if the number of participating students is at least 3.

\begin{figure}[htbp]
   \centering
   \includegraphics[width=5cm]{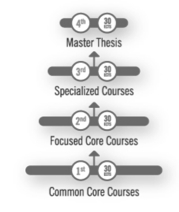} 
   \caption{The workflow process}
   \label{fig:archi2}
\end{figure}

We have designed a study program that is organised into four semesters, each 30 points worth (Figure \ref{fig:archi2}). It consists of an orientation semester with common core courses, followed by the second semester with Focused Core Courses, the specialisation semester (with specialisation courses), and the Master Thesis (fourth semester). This architecture has been chosen as the transfer of knowledge regarding an academic education affords time and training. We believe that this is less guaranteed in two semesters only.

\subsection{Structure}\label{structure}

A second aspect has been to direct the research orientation instead of focusing on a too wide orientation. Matching this up with existing knowledge strengths and forces, this has finally led to the specialisations given in Figure \ref{fig:archi1}. The idea of offering five specialization but mastering them with the existing sources has been solved by synchronising courses in the first and second semester (section \ref{courses}).

Each course receives a \textsf{letter mark}. A \textsf{letter mark} indicates to which specialisation a course belongs to. In this respect and since all courses of the first semester are part of a student's self-orientation, they receive both an \textsf{A} for Intelligent and Adaptive Systems, a \textsf{B} for Bio-informatics, a \textsf{C} for Communicative Systems, an \textsf{E} for Advanced Software System, and a \textsf{S} for Security and Trust. In the second semester, courses are only marked with such specialisation symbols to which they belong to. All courses are 5 ECTS worth. A student must follow this classification in a way that (s)he select four courses of her/his specialisation plus two courses that are not labelled with the selected specialisation letter. Depending on the student's selection he may stay double-tracked, meeting the conditions for two specialisation simultaneously. In the third semester, independent courses are offered that uniquely are assigned to one specialisation. Each course is 4 ECTS worth, a student must visit 4 of them, plus 2 courses from another specialisation. The courses \textsf{Intellectual Property} and \textsf{Project Management} are mandatory to all students and are 3 ECTS worth. The last semester is fully dedicated to the Master Thesis, which takes place in the chosen specialisation and which counts 30 ECTS (section \ref{master}). As the program is research-oriented, the Master thesis takes place in the scope of a research project. Alternatively, the Master Thesis can be done in co-operation with industry.

\subsection{Specializations}\label{specialization}
\begin{figure}[htbp]
   \centering
   \includegraphics[width=8cm]{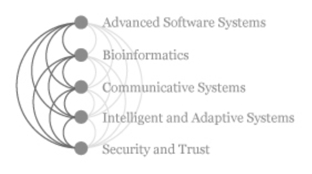} 
   \caption{List of specialisation, with Intelligent and Adaptive Systems (\textsf{A}), Bio-informatics (\textsf{B}), Communicative Systems (\textsf{C}), Software Engineering (\textsf{E}), and Security \& Trust (\textsf{S}).}
   \label{fig:archi1}
\end{figure}

Once, a student has chosen her/his specialisation, (s)he may switch once to another specialisation. For this, the student must collect all the ECTS from his new specialisation. He looses ECTS, if the ancient visited courses do not fit into the new course system. In Figure \ref{fig:archi1}, the different specialisation are listed; this list compounds typical computer science fields and bio-informatics, which is performed in co-operation with the Dept. of Biology.

The Master in Information and Computer Sciences program has to be finished by choosing one supervisor and a Master Thesis of the student's interest. A Master Thesis is a major piece of original work that includes research; it defines a formal written description of that research, and an oral defence of the research as well. A Master Thesis should contribute new knowledge to the specified discipline, but will include an extensive review of what others have contributed to the topic as well. Overall, the tone should be scholarly, with a primary audience of other information science researchers. The Master Thesis is of a more complex nature than a term paper but less than a doctoral dissertation. For example, a synthesis and description of others' research and writings alone may be appropriate as a term paper. Such a comprehensive \textsf{review of the literature} must be included as a subsection of the thesis but is not sufficient by itself. On the other hand, a doctoral dissertation might use an experimental or survey methodology involving large numbers of subjects on a national or international level. A Master Thesis may be narrower in scope, being restricted to a local problem or a smaller number of subjects. A Master Thesis should not just be an expression of an opinion; conclusions and opinions must be based on research results and analysis.

\subsection{Example}\label{example}

Assume, the following courses $c^{Semester}_{ListOfSpec}$ are offered where $Semester$ corresponds to the semester number and $ListOfSpec$ to the list of specialization, indicated by the corresponding letter mark:

\begin{itemize}
\item[{$\circ$}] First Semester courses $\sigma_1$ = ($c^1_{ABCES}$, $c^1_{ABCES}$, $c^1_{ABCES}$, $c^1_{ABCES}$, $c^1_{ABCES}$), 30 ECTS worth.
\item[{$\circ$}] second semester courses $\sigma_2$ $\sigma_2$ = ($c^2_{B}$, $c^2_{B}$, $c^2_{A}$, $c^2_{C,E}$, $c^2_{C,S}$, $c^2_{A,C,S}$, $c^2_{A,B}$, $c^2_{A,B}$, $c^2_{C,S}$, $c^2_{S}$, $c^2_{E}$, $c^2_{E}$, $c^2_{E}$) (each 5 ECTS worth).
\item[{$\circ$}] third semester courses $\sigma_3$ $\sigma^1_3$ = $c^3_{A_1},\dots, c^3_{A_4}$, $c^3_{B_1},\dots, c^3_{B_4}$, $c^3_{C_1},\dots, c^3_{C_4}$, $c^3_{E_1},\dots, c^3_{C_4}$, $c^3_{S_1},\dots, c^3_{S_4}$), each 4 ECTS worth, and $\sigma^2_3$ = ($c^3_{X}, c^3_{Y}$), each 3 ECTS worth.
 \end{itemize}

Then a student chooses all courses in the first semester, all courses labelled with an \textsf{A} if (s)he chooses specialisation A plus two other courses, for example $c^2_{B}$ and $c^2_{B}$, in the second semester. As the student visits the requested four courses both for specialization \textsf{A} and \textsf{B}, (s)he may select between these specialization right after the second semester. In the third semester, the student visits the four courses from the chosen specialisation (either \textsf{A} or \textsf{B}) and two other courses in $\sigma^1_3$  plus the two courses from $\sigma^2_3$.

\section{Implementation}\label{implementation}

\textsf{MICS} is geared towards mastering state-of-the-art technologies and emerging technologies and prepares participants for research in specific domains. The program textsfasises the importance of a long-term perspective and recognizes the need for solid theoretical foundations to pave the way for lifelong learning. The \textsf{MICS} program aims to provide a modern curriculum in the spirit of the Bologna agreement (\cite{bologna}), focusing on five core areas. In the opinion of the review team, the profile and the goals of the program are appropriate to a modern high-level program in the information and computer sciences.

The \textsf{MICS} program connects closely with the research areas (sections \ref{labs}, \ref{p1}) offered within the faculties respectively the departments that are responsible for the program. This should provide an excellent environment for the education and training offered in the program. The young researchers and professionals that are the progamme's target group represent a strategically essential asset for the information society. In this sense, the program is well-positioned, with sound scientifically ambitions in its five core areas. Besides the overall profile of the program, the profile of the core areas of the curriculum needs to be clearly communicated as well, including the competences which the students are expected to acquire by successfully completing a core area.

\subsection{Semester Organisation}\label{semester}
The academic year is split into a winter semester and a summer semester. The winter semester regularly starts in the mid of September and ends in the mid of February. The summer semester starts in the mid of February and ends in the mid of September. The lectures take place between September and December (winter), and between February and May (summer), respectively. The lecture time is followed by one preparation week and an examination period, which takes three weeks.

\subsection{Distributing the responsibilities}\label{responsibilities}
An important aspect in architecting \textsf{MICS} is the distribution of tasks in general, instead of concentrating the work to one person or to a small group. With this, certain \textsf{MICS} responsibilities can be done in parallel and can be temporarily taken over by someone else. With this, we have established a certain number of. These are inside MICS

\begin{itemize}
   \item[{$\circ$}] the \textsf{director of studies}
   \item[{$\circ$}] the Vice-\textsf{director}
   \item[{$\circ$}] Specialisation head, one for each specialisation
   \item[{$\circ$}] the \textsf{MICS} secretariat
   \item[{$\circ$}] an independent \textsf{MICS} Jury, consisting of the director and the vice-director, the specialization heads, and the secretariat.
   \item[{$\circ$}] the board of tutors
   \item[{$\circ$}] one responsible person for Printing Services
\end{itemize}

and outside MICS

\begin{itemize}
   \item[{$\circ$}] a responsible contact person for \textsf{MICS} Marketing, directly working with the Marketing Department of the University
   \item[{$\circ$}] a contact person for the \textsf{MICS} Enrollment  (section \ref{residence})
   \item[{$\circ$}] one responsible person at SEVE
   \item[{$\circ$}] one responsible person for cultural activities (section \ref{sports})
   \item[{$\circ$}] one responsible person for housing (section \ref{housing})
   \item[{$\circ$}] one responsible person for job portal (section \ref{ack})
\end{itemize}

The \textsf{director of studies} co-ordinates \textsf{MICS} in general and is supported by the vice-rector and the \textsf{MICS} secretariat. Beside of that, the \textsf{director} and vice-\textsf{director} coordinate the first and second semester courses $\sigma_2$ of \textsf{MICS}. The specialisation heads are responsible for the third semester courses $\sigma_3$ and the association of the Master Thesis. In the first case, they organise at least four courses of their specialisation, for the last case they decide if the intended thesis fits into the specialization. The independent \textsf{MICS} jury meets twice per month and discusses related issues regarding the students, their performances, and eventually necessary changes. Additionally, they advise a student to participate in a tutoring session or draw any consequences in case the students'result do not promise a successful progress.

The enrolment is completely transferred to the student service department  (\cite{application}, \cite{contact}, \cite{seve}), respectively to the \textsf{MICS} representative at the student service department. They have established an online questionnaire, which is available in the english language. After having filled out this form, the student send the application documents to the service department. The student service is also responsible for the mobility of students (\cite{mobility}).

The \textsf{MICS} marketing is performed through a liaison-officer who stays directly in contact with the Marketing Department of the University. He is supported by the the Web Design responsible. Other \textsf{MICS} services are performed by the responsible for student housing (\cite{logement}) at the University and a local person who organizes social activities. A job portal is offered online (\cite{jobportal}), cultural activities are offered and supported (\cite{sports}).

\subsection{Enrolment and Numerus Clausus}\label{numerus}

The following requirements have to be met to be able to successfully complete the studies. First, the students must be able to read English with ease and to understand spoken English in lectures and group discussions. The student has to prove

\begin{itemize}
  \item[{$\circ$}] the ability to understand spoken English (listening Comprehension).
  \item[{$\circ$}] to read and understand both technical and non-technical material (reading Comprehension).
  \item[{$\circ$}] to write standard written English (writing skills).
\end{itemize}

and must be fluency in speaking. In case the student has not demonstrated his expertise of the English language, we recommend an intensive full-time course in English before entering the first semester. Throughout the first semester, the student will attend an English course concerning specific terms related to computer science. The student needs to have a basic education in computer science as given in a Bachelor of Science in Computer Science. This includes e.g. mathematical foundations as given in (Linear) Algebra, Analysis, etc.,  a profound programming skills in at least one programming language comprehension of abstract subjects, and a basic understanding of computer science.

There is no need for work experience, but it can become a plus for a student to share industrial experiences. The study is aimed at a continuation within doctoral studies. Hence, interest in research-oriented methods is required. In concern of the marks, we have implemented a Numerus Clausus of at least a \textsf{bien/15} (french system), \textsf{gut/2.3} (german system), or \textsf{good/B-} (american system).

\subsection{Procedure of selecting the Candidates}\label{selecting}
Applications to \textsf{MICS} are solely allowed between April and September as \textsf{MICS} only starts in the winter term. Within this time frame, the procedure of selecting the candidate is done regularly twice per month. After having filled out the online form and sent the application documents to the student service department, the application dossier is sent to the \textsf{director of studies}. Then, the \textsf{director} decides in collaboration with the Vice-\textsf{director} and the specialization heads that the candidate is either accepted, rejected, or that currently no decision can be made. In the last case, some more information about the candidate is requested. The selection procedure is done twice per month. 

\subsection{Tutoring}\label{tutoring}
Every student can choose a professor as the contact person of his trust. This person advises and consults the student in study-relevant aspects and supports him/her in social conflicts. The contact person is not responsible for scholarships and things that concern her/his stay in Luxembourg.

For example, a candidate may select another specialization once. He/she resigns from a selected specialization and joins a new specialization. The Master Thesis can then be done in this specialization. If the candidate selects a new specialization, then he/she has to ensure that he/she has collected sufficient ECTS following this new specialization. The candidate can transfer received ECTS from the old specialization to the new selected specialization, if a course is assigned to this new selected specialization.

\subsection{Courses}\label{courses}

Courses of the first semester ($\sigma_1$), second semester ($\sigma_2$), and two courses from the third semester (Project management, Intellectual Property) are fix, whereas the specialisation courses of the third semester ($\sigma^1_3$) may change - depending on students' interest and existing trends. The program starts with an \textsf{orientation week} where all new students may get to know to fellow students, doctoral students, and professors.

\subsubsection{Orientation - Core Courses $\sigma_1$}\label{core}

The Core Courses $\sigma_1$ are compulsory for all students, they act as an orientation for the student. Four courses plus one non-technical course must be visited by every student. 

\begin{itemize}
  \item[{$\circ$}] Theoretical Foudations
  \item[{$\circ$}] Communication and Networking
  \item[{$\circ$}] Distributed Systems
  \item[{$\circ$}] Intelligent Systems
  \item[{$\circ$}] Technical English
\end{itemize}

The course on Theoretical Foundations is concerned with several mathematical preconditions that are important for the further studying. The other technical courses are concerned with basic aspects in computer science that are further relevant throughout the studies. The non-technical lesson (Technical English) supports students to improve their english language skills in the field of information and computer sciences and to guide them in reading scientific articles.

\subsubsection{\textsf{A} - Second semester $\sigma_2$ and third semester courses $\sigma^1_3$}\label{a}

The area Intelligent and Adaptive Systems is concerned with computing in complex and dynamic environments given limited resources and incomplete or uncertain knowledge. We investigate the theoretical foundations and the algorithmic realization of systems performing complex problem solving with a high degree of autonomy (intelligent), and exploiting learning to deal with opaque and changing contexts (adaptive). The major areas of interest are

\begin{itemize}
   \item[{$\circ$}] Evolutionary Computing, Agent Systems, Distributed Coordination
   \item[{$\circ$}] Data/Text Mining, Knowledge Discovery, Machine Learning
   \item[{$\circ$}] Knowledge Representation, Information Management
   \item[{$\circ$}] Information Theory, Uncertain Inference
\end{itemize}

The second semester $\sigma_2(A)$ and third semester courses $\sigma^1_3 (A)$ are then:

\begin{itemize}
   \item[{$\circ$}] Evolutionary Computing
   \item[{$\circ$}] Information Theory and Coding
   \item[{$\circ$}] Knowledge Discovery and Data Mining
   \item[{$\circ$}] Knowledge Representation
   \item[{$\circ$}] Applied Mining in Security
   \item[{$\circ$}] Argumentation
   \item[{$\circ$}] Content Management Systems
   \item[{$\circ$}] Natural Language Processing and Text Mining
   \item[{$\circ$}] Selected Topics In Artifical Intelligence
   \item[{$\circ$}] Stochastic Methods in Intrustion Detection
\end{itemize}

\subsubsection{\textsf{C} - Second semester $\sigma_2(C)$ and third semester courses $\sigma^1_3(C)$}\label{c}

The area of \textsf{Communicative Systems} encompasses both communication as such (Information Transfer) as well as distributed systems communicating with each other (Communicating Systems). Communication networks are built by interconnecting multiple channels, allowing to establish point-to-point communication links, finally between end devices. Knowledge of both the theory behind as well as practical aspects is essential. For instance information routing, distribution strategies and reliability in static, mobile, ad-hoc, peer-to-peer, and grid networks are crucial for the creation, operation, maintenance, and optimization of heterogeneous networks.
In order to properly design robust communicative systems, a thorough understanding of the properties and characteristics of distributed systems in general is vital. Introducing distribution into the design of systems results in much more complex architectures, obeying different semantics compared to centralized approaches, and requires more elaborate techniques for their realization and testing.

The implementation of communicative systems, due to their complexity, is typically relying on sophisticated middleware services as well as application frameworks. Hence, knowledge both in using as well as in designing such software artefacts is essential. In this respect, the major courses take place in the second ($\sigma_2$) and third semester ($\sigma^1_3$):

\begin{itemize}
  \item[{$\circ$}] Mobile Computing
  \item[{$\circ$}] Security in Static and Dynamic Network Layers
  \item[{$\circ$}] Information Theory and Coding
  \item[{$\circ$}] Cryptography
  \item[{$\circ$}] Coding Theory
  \item[{$\circ$}] Non-/Cooperative Information Routing
  \item[{$\circ$}] Parallel and Grid Computing
  \item[{$\circ$}] Technical Systems Modeling and Simulation
  \item[{$\circ$}] Ubiquitous Computing
\end{itemize}

\subsubsection{\textsf{E} - Second semester $\sigma_2(E)$ and third semester courses $\sigma^1_3(E)$}\label{e}

\begin{itemize}
   \item[{$\circ$}] Mobile Computing
   \item[{$\circ$}] Dependable Real-time Systems
   \item[{$\circ$}] Formal Methods
   \item[{$\circ$}] Evolutionary Computing
   \item[{$\circ$}] Product Lines Engineering
   \item[{$\circ$}] Model-Driven Software Development
   \item[{$\circ$}] Proactive Systems
   \item[{$\circ$}] Research Frontiers
   \item[{$\circ$}] Service-Oriented Software Architecture
   \item[{$\circ$}] Embedded Systems
\end{itemize}

\subsubsection{\textsf{S} - Second semester $\sigma_2(S)$ and third semester courses $\sigma^1_3(S)$}\label{s}

Security (of information systems) refers to the fact that protection goals are achieved in spite of intelligent attacks. Some important protection goals are :

\begin{itemize}
   \item[{$\circ$}] confidentiality, i.e. avoid an unauthorized acquisition of information
   \item[{$\circ$}] integrity, i.e. avoid unauthorized modification of information
   \item[{$\circ$}] availability, i.e. avoid unauthorized restriction of the functionality of information systems
   \item[{$\circ$}] accountability, i.e. be able to identify a responsible person for each transaction.
\end{itemize}

In open communication systems (e.g. the Internet or mobile communication systems) it is unrealistic to assume that all parties involved completely trust one another, since all parties have to be regarded as potential attackers. Trust, therefore, is the belief in the behaviour of an involved party for some given purpose. It is closely bound to human nature and to the quality of the security available. The second semester courses $\sigma_2$ are

\begin{itemize}
   \item[{$\circ$}] Application of Trust Systems
   \item[{$\circ$}] Cryptography
   \item[{$\circ$}] Information Theory and Coding
   \item[{$\circ$}] Security in Static and Dynamic Network Layers
   \item[{$\circ$}] Advanced Open Network Security
   \item[{$\circ$}] Applied Mining in Security
   \item[{$\circ$}] Cryptography in the real world
   \item[{$\circ$}] Management of Information Systems Security
   \item[{$\circ$}] Verification of Security Protocols
\end{itemize}

\subsubsection{About the Master Thesis}\label{master}

A Master Thesis in \textsf{MICS} is a major piece of original work that includes research; it defines a formal written description of that research, and an oral defence of the research as well. A Master Thesis should contribute new knowledge to the specified discipline, but will include an extensive review of what others have contributed to the topic as well. Overall, the tone should be scholarly, with a primary audience of other information science researchers. The Master Thesis is of a more complex nature than a term paper but less than a doctoral dissertation. For example, a synthesis and description of others'research and writings alone may be appropriate as a term paper. Such a comprehensive review of the literature must be included as a subsection of the thesis but is not sufficient by itself. On the other hand, a doctoral dissertation might use an experimental or survey methodology involving large numbers of subjects on a national or international level. A Master Thesis may be narrower in scope, being restricted to a local problem or a smaller number of subjects. A Master Thesis should not just be an expression of an opinion; conclusions and opinions must be based on research results and analysis. 

A candidate must perform the Master Thesis in a specialization that he/she has selected. The Master Thesis is supervised by a professor of the department of the University of Luxembourg. This person is called first supervisor. The first supervisor may be supported by a Postdoctoral or a doctoral student. 

After having submitted the Master Thesis, the \textsf{director of studies} selects a second supervisor (second supervisor). The Master Thesis process starts with the opening and ends with the closing of the Master Thesis. The candidate starts the Master Thesis process either in the third or fourth semester. The Master Thesis process is opened by the candidate. He/she must include: the start date, the working title of the Master Thesis, the selected specialization, and the name of the first supervisor. The working title may change. The starting date is mandatory. The Master Thesis compounds a written documentation and a defence. The candidate has to deliver three printed exemplars of the Master Thesis to the \textsf{MICS} secretariat: one exemplar is sent to the first supervisor, one to the second supervisor, and one to the CSC library. The candidate has to sign a declaration of authorship where he assures that (s)he has written the Thesis by his/her own. The Master Thesis must be written in English language.

The Master Thesis is accepted if the final mark is at least a 10 (following the R\'eglement Grand-Ducal). In this case, the candidate receives 30 ECTS. The Master Thesis process is then closed. The Master Thesis documentation is evaluated by the first supervisor and the second supervisor. Ideally, both should come from the same specialization, but the second reviewer may also share another specialization. This is to be decided by the \textsf{director of studies}. The second supervisor may come from outside (industry, university, research institution, etc.). (S)he then must have at least a doctorate, which is to be decided by the \textsf{director of studies}.

The Master Thesis must be presented by the candidate within a defence, which includes a 30 minutes presentation and the answering of questions within a discussion that may not exceed one hour. The defence is a public session, it is headed by a \textsf{defence board} that consists of the \textsf{director of studies}, the \textsf{specialization head} of the specialization where the Master Thesis belongs to, the \textsf{second reviewer}, and up to two other persons of the specialization. The marks follow the classification of the R\'eglement Grand-Ducal; the final mark of the Master Thesis is then calculated as follows:

\begin{itemize}
   \item[{$\circ$}] weight of the first supervisor: 40\%
   \item[{$\circ$}] weight of the second supervisor: 20\%
   \item[{$\circ$}] weight of the defence board: 40\%
\end{itemize}

A Master Thesis can be repeated once. If a Master Thesis is repeated, then the candidate may change the specialization. If a Master Thesis is repeated, then the content should be significantly different to the previous Master Thesis. The Master Thesis may be performed in cooperation with another university, for example in the scope of EU programs or through individual cooperations with universities or research institutions. The Master Thesis can be done in cooperation with an industry, if the content of the Master Thesis follows a scientific nature. This is to be decided by the first supervisor, the specialization head and the \textsf{MICS} \textsf{director of studies} (equal vote). In case that a decision can not be found, the specialization head has to decide.

\subsubsection{Course descriptions}\label{description}
The program of \textsf{MICS} is published in the web (\cite{home}). Furthermore, information to all offered courses are available on individual pages. Each page contains a description of the course content, the aims of qualifications/outcome, the pre-conditions, and the name of the responsible persons. Depending on the lecturer, course material and references can be downloaded.

The teaching methods are chosen by the lecturer. A lecturer is either a professor or a person that owns a doctorate. The course may be supported by a PhD student or a Master student of a higher semester. A course can be held week by week or as a block event. The lecturer decides how a course is performed. This may be a mixture of a lecture,  an exercise, a seminar, a colloquium, and a practical project.

Depending on the nature of the course, the examination may be oral or written examinations. Each course result and the Master Thesis yields on a mark that bears on the directives of the \textsf{R\'eglement Grand-Ducal}. All examinations can be taken every semester. 

Examinations of the first, second, and third semester can be performed either oral or written depending on the lecturer twice a year. Written examinations in the first, second, and third semester take at least one hour and at most two hours. Oral examinations the first and second semester take at least thirty minutes. Oral examinations in the third semester takes up to one hour. All written examinations are performed in English. Oral presentations can be performed either in English or in another language if both parties agree. If there is no agreement, then the examination will be done in English.

The marks and grades are given following the regulations and directives of the \textsf{R\'eglement Grand-Ducal}. We follow the french system, meaning that an \textsf{excellent} corresponds to a mark between 18 and 20, a \textsf{very good} to a mark between a 16 and 17.9, a \textsf{good} to a mark between 14 and 15.9, a \textsf{satisfactory} to a mark between 12 and 13.9, and a \textsf{sufficient} to a mark between 10 and 11.9. If an examination is not passed, it is failed (less than 10).

\subsection{Cooperation and Mobility}\label{mobility}

There exist a certain number of cooperations, which are continuously expanded. Currently, cooperations with the following partners are given:

\begin{itemize}
   \item[{$\circ$}] University-to-University co-operation agreement with the University of Shadong, China.
   \item[{$\circ$}] Socrates Agreement with University of Malaga, Spain, and the University of Eindhoven, The Netherlands.
   \item[{$\circ$}] Diverse individual cooperations (on department level) and personal relationships. 
    \end{itemize}

Furthermore, we are still in preparing an Erasmus Mundus agreement with the University of Nancy, France, University of Udine, Italy, and the Technical University of Munich, Germany.

\subsection{CSC Research Labs}\label{labs}
\textsf{MICS} is mainly performed by faculty members from the Computer Science and Communications Research Unit (CSC, see \cite{csc}). \textsf{MICS} students can perform their master theses in collaboration with the CSC research laboratories:

\begin{itemize}
   \item[{$\circ$}] Communicative Systems Laboratory (\textsf{ComSys})
   \item[{$\circ$}] Interdisciplinary Lab for Intelligent and Adaptive Systems (\textsf{ILIAS})
   \item[{$\circ$}] Laboratory of Algorithmics, Cryptology and Security (\textsf{LACS})
   \item[{$\circ$}] Laboratory for Advanced Software Systems (\textsf{LASSY})
\end{itemize}

\subsection{MICS Facilities}\label{library}
For \textsf{MICS}, both a computer room and a library is available. The computer room encompasses 20 computers, the library a compound list of books relevant to the studies.

\section{Partners outside \textsf{MICS}}\label{external}

\subsection{Centre of Culture and sports}\label{sports}

A cultural program is provided as well as a call for participation in diverse sport activities (\cite{sports}).

\subsection{Centre of student mobility}\label{mobility}

A centre of student mobility is established by the University of Luxembourg (\cite{mobility}).

\subsection{Centre of student accomodation}\label{housing}

The University offers several student houses and apartments (\cite{students}).

\subsection{The Job Portal}\label{jobportal}

\begin{figure}[htbp]
   \centering
   \includegraphics[width=12cm]{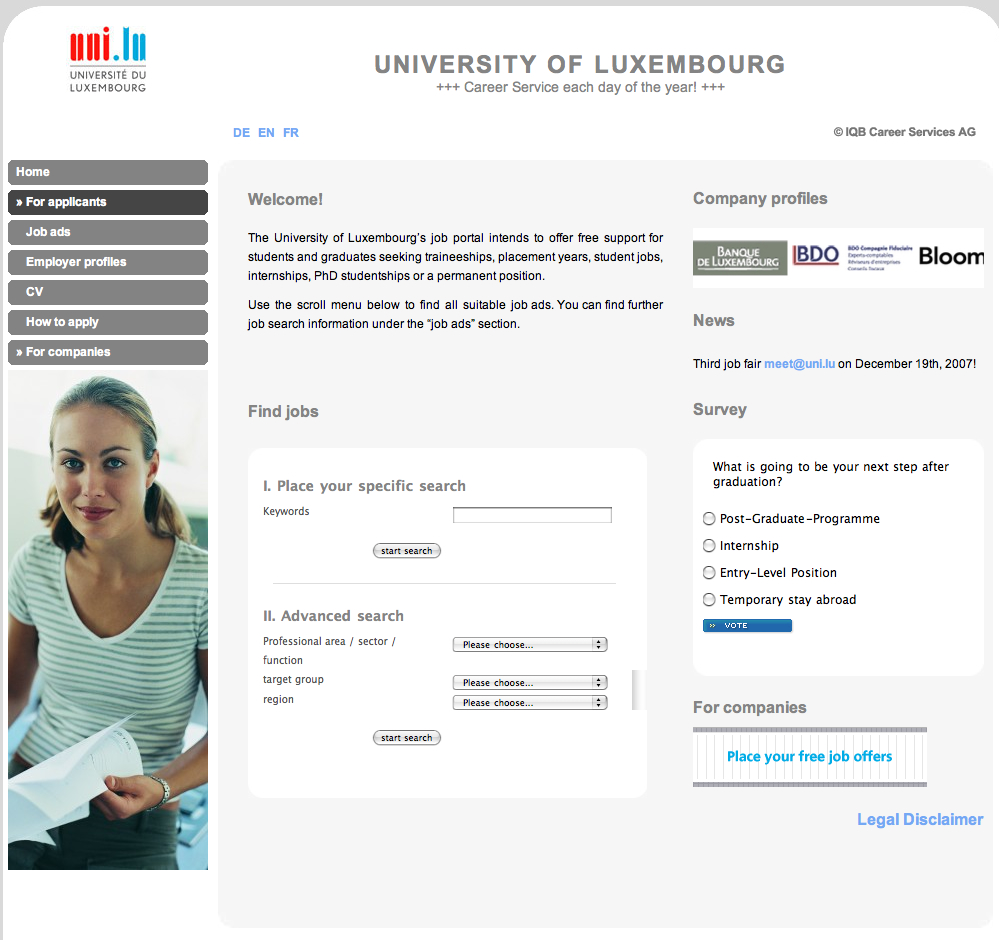} 
   \caption{The Job Portal at the University of Luxembourg}
   \label{fig:portal}
\end{figure}

The University of Luxembourg's job portal intends to offer free support for students and graduates seeking traineeships placement years, student jobs, internships, PhD studentships or a permanent position (Figure \ref{fig:portal}).

\subsection{Entry and Residence}\label{residence}
No special formalities\footnote{Status: April 2008} are required for students who are citizen or residents of a member state of the European Union. A student has to make an entry declaration to the community administration in which they are resident. For students from a non-EU state, the following procedure must be followed to enter luxemburgish territory:

\begin{itemize}
     \item[{$\circ$}] The student has to go to the embassy of Luxembourg and/or of Belgium in the student's home country. The student presents there a certificate of pre-inscription to obtain your visa.
     \item[{$\circ$}] As soon as the student arrives on the territory of Luxembourg, (s)he has to contact the \textsf{Ministry of Foreign Affairs} with the following documents:
      \begin{itemize}
         \item[{$\circ$}] a certificate of residence
         \item[{$\circ$}] a medical certificate obtained by a doctor in Luxembourg
         \item[{$\circ$}] a deposit of guarantee amounting to 1200 Euro in a bank in Luxembourg
         \item[{$\circ$}] a pre-inscription document at the University of Luxembourg
       \end{itemize}
\end{itemize}
       
More information can be received from the \textsf{Minist\`ere des Affaires \'etrang\`eres et de l'Immigration}: 5, rue Notre-Dame; L-2240 Luxembourg. Phone: +352 478 23 00, Fax: +352 22 31 44.

\subsection{Financial Support}\label{financial}

The state financial aid\footnote{Status: April 2008} is a financial contribution that the Luxembourg State gives to students who take a higher education (amended law of 22 June 2000, see \cite{financial}). The financial aid can consist of two parts:

\begin{itemize}
     \item[{$\circ$}] The grant that is paid by the state into the student's personal account.
     \item[{$\circ$}] The loan is set to an interest rate of only 2\% with one of the banks that have signed an agreement with the State: Banque et Caisse d'Epargne de l'Etat, Banque G\'en\'erale du Luxembourg, Banque de Luxembourg,  Banque Raiffeisen, Banque ING, DEXIA-BIL, and Fortuna Banque.
\end{itemize}

The re-payment conditions are then that the student must start the repayment two years after the study has finished or stopped. From then onwards, the loan must be repaid within 10 years, except for students who are more than 35 years old (5 years). The 2\% interest that is payable by the student is regularly debited twice per year.

All students can benefit from the financial aid if they are registered in a course of post-secondary, higher or university studies and who comply with the following conditions of nationality to be a
\begin{itemize}
     \item[{$\circ$}] Luxembourg national and to live in Luxembourg
     \item[{$\circ$}] national of a European Union member country, to live in Luxembourg and to come within the scope of EEC regulation No 1612/68 regarding the free circulation of workers within the Community.
     \item[{$\circ$}] national of third party States or a stateless person, living in Luxembourg for at least 5 years and holding a secondary education diploma recognised as being equivalent by the Ministry of National Education,
     \item[{$\circ$}] political refugee in Luxembourg and to be living in Luxembourg.
\end{itemize}

To obtain the financial aid, a student must complete and confirm online (\cite{cedies}). It is also possible to send a written application to \textsf{CEDIES}, 211, route d'Esch, L-1471 Luxembourg, Fax:  45 56 56  EMail: cedies@mcesr.etat.lu, and to send back the written questionnaire completed and supplemented by the required documents to the same address, meeting the deadlines for this. More information can be found at \cite{cedies} and \cite{financial}.

\section{Results}\label{results}

\subsection{Some Statistics}\label{stats}
The number of professors has become doubled in the last four years, as two departments of two faculties have merged and new professors hired. 25\% of the professors come from Germany and Luxembourg, 10\% from the Netherlands and France, respectively. 15\% are originally from Belgium and 5\% from Sweden, Austria, and Russia, respectively. The average age is lower than 45, the common language is English. Some external experts, for example native English speaker from the UK and Ireland, are on board as well. Currently, an interdisciplinay center will become established, with a director to be involved in \textsf{MICS} as well.

In concern of applications and the acceptance rate of students within the last years, the number of interested candidates filling out the contact form vs. the number of students who have submitted their documents widely diverges. For example, the number of interested students in 2007 had been over 200 whereas the number of submitted dossiers had been lower than 50. In general, the acceptance rate of students has been swung into 30-35\%. We do not have an intended maximum number of students per year, but accept all students who satisfy the entry requirements.

In respect to the nationalities of students, there exist no trend where the students come from. Historically, diverse African students still benefit from a governmental agreement in receiving financial support. A co-operation with the Shadong University of China have been initiated, and the participation in Erasmus Mundus and Socrates programs of the European Union are still supported. One might think that the students from Luxembourg itself intend to study: this is mostly not true as the long-establihsed universities abroad offer a more attractive program to many Luxemburgers. 

Currently, the nationalities are distributed as follows:
\begin{center}
\begin{tabular}[ht]{|c|c|}
 \hline
  Continent & Number (in \%)\\  \hline
  Europe & 63\\
  Africa & 17\\
  Asia & 16\\
  South-America & 4\\ \hline
 \end{tabular}
 \label{fig:example}
\end{center}

with european students are most from Luxembourg (45\%) but also from Spain (20\%), Germany (10\%), and Belgium, Greece, Turkey, and UK (5\% each).

There exist a couple of reasons why accepted students leave the program. First, many students submit their proposal to other universities as well, and, being accepted, they choose on studying there. Second, students misuse the acceptance to \textsf{MICS} to bridge their time to their industrial job. Third, luxembourgish students inscribe because of fiscal reasons. Fourth, \textsf{MICS} is too difficult as the program does not meet the student's expectations or that the requirements are too high-leveled. Students therefore decide either themselves to stop the program or are advised by the \textsf{MICS} jury to think about exiting. Fifth, students find a job in Luxembourg. Sixth, students stay only temporarily at \textsf{MICS}, because they are member of European programs. Sixth, some students do not start because of financial problems. Seventh, students do nor (or too late) their permit to stay (non-EU).

So far, the following table shows the number of students who have left the master program in the last 4 years.

\begin{center}
\begin{tabular}[ht]{|c|c|}
 \hline
  Reason & Students have left (in \%)\\ \hline
  1 & 25\\
  2 & 9\\
  3 & 12\\
  4 & 6\\
  5 & 6\\
  6 & 36\\
  7 & 6\\ \hline
 \end{tabular}
 \label{fig:example}
\end{center}

Concerning the gender in general, the number of male students is less higher than female students. Whereas the number of male students who have already finished their studies is six times higher than female students, the current number of female and male students are identical.

33\% of the students have selected the specialization \textsf{A}, 43\% the specialization \textsf{C}, and 24\% the specialization \textsf{S}. 

\subsection{Quality Assurance through Accreditation}\label{qa}

The accreditation has initiated by the Computer Science Department one year before \textsf{MICS} has start, thus in the mid of 2004. The accreditation has kindly be supported by our former vice-rector Dr. Adelheid Ehmke.

The accreditation process has been performed by the accreditation institute AQAS, located in Bonn, Germany. A liaison-officer has been defined who continuously has acted as the contact person. He has led us through the process while telling us the guidelines, keeping us informed about the work of the accreditation committee, and organizing the evaluation days. The accreditation committee has been an international group consisting of professors, student and industrial representative.

The inspection has taken place in June 2005 over a period of three days. Beside of a survey regarding the infrastructure and an intensive discussion with the \textsf{MICS} management, the following discussions with existing Bachelor students, the \textsf{MICS} staff and the rectorate of the University has taken place. The general outcome has been very positive, but there has been a certain number of criteria to be fulfilled. Therefore, based on the reviewer's report and consultations during the 20$^{th}$ session of the accreditation commission on October 10th, 2005, the accreditation commission has issued a \textsf{conditional accreditation} for the research-oriented Master program \textsf{MICS} offered by the University of Luxembourg. The accreditation had been valid until the end of September 2009. The fulfillment of the conditions has documented and submitted to the commission for approval by September 2006. After a re-evaluation, the conditional accreditation has then been changed to a \textsf{full accreditation}. Moreover, a certain number of recommendations have been given, for example that a diploma supplement must be given, that the result of a standardized English test (like TOEFL) should be used as to determine if a student's language skills qualify for entry to the program, and that decisions on which specialization could be offered should be made known at the earliest possible date (otherwise student expectations might be disappointed). Also, documents relating to the Master's program should be available in English and that the quality management concept being under development at the university level and to be in place within one year, should become public. The commission has said that a catalogue should be drawn up defining responsibilities (and competencies), quality objectives, quality indicators, and areas of quality measurement (e.g. external evaluation, alumni). The commission has wished to stress that access to electronic libraries should be provided at an early stage. The accreditation commission wishes to stress the importance of maintaining a website for the program and creating an intranet, as suggested by the department.

Students are required to have a good command of the English language as courses are instructed in English, There is no need for work experience, but it may be advantageous for students to share their respective industrial experience with each other. The faculty states that in a world characterized by increasing competition for minds, Luxembourg needs well-educated specialists who are able to deal with the 21$^{st}$ century challenges in information technology. Of particular interest to the university are those areas relevant to the local finance and media industry, or the large administrative sector, as well as which which reinforce existing research strength within the university, notably
security and trust systems, advanced telecommunication systems, and large-scale processing of data, information, and knowledge as well as engineering of complex software-intensive systems and computational techniques for the life sciences.

\subsection{Establishing Research Groups}\label{rgroups}
There are a certain number of results that has come up with \textsf{MICS}.  Most importantly, research had been consolidated in four main directions. The identification process or self-discovery had been the major side-effect.

\subsubsection{ComSys}\label{comsys}
The Communicative Systems Laboratory (\textsf{ComSys}) is part of the Computer Science and Communication Research (\cite{csc}) and focuses on state of the art research in digital communications. Embracing the end-to-end arguments in system design, ComSys focuses on integrated research in the areas of Information Transfer and Communicating Systems. Information Transfer is concerned with information transmission over potentially complex channels and networks. Communicating Systems in turn are the composition of multiple distributed entities employing communication networks to collaboratively achieve a common goal. ComSys has strong technical and personal facilities to improve existing and develop new solutions in the following research topics:

\begin{itemize}
    \item[{$\circ$}] Information Transmission
    \item[{$\circ$}] Wireless Communication Systems
    \item[{$\circ$}] Security Protocols
    \item[{$\circ$}] Trust Models
    \item[{$\circ$}] Middleware
    \item[{$\circ$}] Parallel and Distributed Systems
    \item[{$\circ$}] Grid and Peer-to-Peer Computing
    \item[{$\circ$}] Management and Mining of Data
\end{itemize}

The research fields will have a strong impact on the 21$^{st}$ century. The rapidly growing demand for information exchange in people's daily lives requires technologies like ubiquitous and pervasive computing to meet the expectations of the information society and novel adaptive concepts tackling the continuing data challenges. The resulting problems have already been a key enabler for some industrial and governmental founded projects at national and European level.

\subsubsection{ILIAS}\label{ilias}

The Interdisciplinary Lab for Intelligent and Adaptive Systems (\textsf{ILIAS}) is a research group within the Computer Science and Communication Research Unit. Its main goal is to realize and develop research and teaching in the area of intelligent and adaptive systems at the University of Luxembourg (and beyond). The overarching subject is information processing - or more specifically, inference - in complex and dynamic environments given limited resources and incomplete or uncertain knowledge. We investigate the theoretical foundations and the algorithmic realizations of systems:

\begin{itemize}
      \item[{$\circ$}] performing complex problem solving with a high degree of autonomy, i.e. intelligent.
      \item[{$\circ$}] exploiting learning to deal with opaque and dynamic contexts, i.e. adaptive.
\end{itemize}

Our research lab is characterized by a strong cross-disciplinary perspective, combining expertise from computer science, information theory, statistics, mathematics, and logic. The principal research directions are:
\begin{itemize}
      \item[{$\circ$}] Global Optimization and Parallel Computing
      \item[{$\circ$}] Information Management and Knowledge Discovery
      \item[{$\circ$}] Information Theory and Stochastic Inference
      \item[{$\circ$}] Individual and Collective Reasoning
      \item[{$\circ$}] Decision Systems
      \item[{$\circ$}] E-learning and Proactive Computing
\end{itemize}

All these areas provide relevant theoretical and practical tools for Security and Trust research and for Bioinformatics.

\subsubsection{LASSY}\label{lassy}

The Laboratory for Advanced Software Systems (\textsf{LASSY}) is conducting research on methods and tools for mastering the development of complex software systems to

\begin{itemize}
      \item[{$\circ$}] develop new engineering processes
      \item[{$\circ$}] investigate modelling languages
      \item[{$\circ$}] perform research on the foundations of software engineering
      \item[{$\circ$}] assist in the development and in the use of e-learning tools
      \item[{$\circ$}] study verification and validation techniques
\end{itemize}

It focuses on application domains like industry-critical systems, e-learning systems, and web-based distributed systems. In 2005, several professors have joined their activities to research areas in
\begin{itemize}
     \item[{$\circ$}] Ambient Systems
     \item[{$\circ$}] Architectures for dependable distributed systems
     \item[{$\circ$}] E-Learning systems
     \item[{$\circ$}] Model-driven engineering
     \item[{$\circ$}] Industry-critical systems
\end{itemize}

\subsubsection{LACS}\label{lacs}
Today, the multimedia technology has expanded to encompass most facets of our daily lives - at work, at school, at home for leisure or learning, and on the move - and it is reaching ever-widening segments of our society. The Internet, e-mails, mobile phones, etc. are already standard channels for the information society to communicate, gain access to new multimedia services, do business or learn new skills. The recent "digital revolution" and widespread access to telecommunication networks have enabled the emergence of e-commerce, and e-government. This proliferation of digital communications and the transition of social interactions into the cyberspace have raised new concerns in terms of security and trust, like: confidentiality, privacy and anonymity; data integrity; protection of intellectual property and digital rights management; threats of corporate espionage, and surveillance system (such as Echelon), etc. These issues are interdisciplinary in their essence, drawing from several fields: algorithmic number theory, cryptography, network security, signal processing, software engineering, legal issues, any many more.

In this context, the research unit LACS which is part of the Computer Science and Communication Research Unit focused on:
\begin{itemize}
    \item[{$\circ$}] Cryptography is the science of protecting secrets. Cryptographic protocols enable to provide secure encryption, digital signatures, and authentication between entities.
   \item[{$\circ$}]  Building a secure cryptographic protocol first requires to clearly specifying the security notions that must be achieved, and then building a protocol that provably achieves these notions.
\end{itemize}

Computational Number Theory is an important tool to build secure public-key cryptosystems. Many proposals for public-key cryptosystems rely on elaborate mathematical objects that are interesting on their own. System and Network Security intends to stop unauthorized users from accessing any part of a computer system. Designing intrusion detection systems will help to determine whether or not someone attempted to break into a system, if he was successful, and what he may have done. Information Security Management includes many topics, like integrity of information, identification of individuals, digital rights management, information risk and policy assessments.

\subsection{Establishing the Research Axis P1}\label{p1}

Today, information technology encompasses many facets of our daily life and it reaches ever-widening segments of our society. New media like the Internet and its services (electronic mailing, World Wide Web, mobile phones) as well as many others have become standard channels for the information society to communicate, gain access to new multimedia services, allow a more effective business and the learning of new skills.

The recent digital revolution and the widespread access to high capacity telecommunication networks have enabled the emergence of e-\{commerce, government\}, and e-science. This proliferation of digital communication and the transition of social interactions into the cyberspace have raised new concerns in terms of security and reliability, like: confidentiality, privacy and anonymity; data integrity; protection of intellectual property and digital rights management; identity theft; threats of corporate espionage, and surveillance system as well as the dependability of complex network systems, the trustworthiness and quality of information sources and services, and many more. These issues are interdisciplinary in their essence, drawing from several fields, being a ground for master students. The main goal of the Security and Reliability priority is to obtain a global visibility, with an impact on the Luxembourg economy. To achieve this goal, three main objectives are identified, namely both the development of a strong internationally recognized CSC wide research center on the area of security and reliability and the development of a regional interface towards security users for Luxemburg and the Grande Region on the area of security and reliability. Additionally, the basis for a world class research institution on security and reliability, for example an interdisciplinary center as defined in the law, or a public-private institution, us provided.

\section{Acknowledgement}\label{ack}
\textsf{MICS} has become prepared and established within more than one year. It has taken efforts to bring it to the level of quality as it is today. In this respect, our greatest thank goes not only to all who have contributed and still participate but additionally to all students who 
give their confidence in our aim to transfer knowledge on a high level.

\section{Conclusions}\label{conclusion}

Today, the \textsf{MICS} program is an accredited master program that has initiated in 2004 and started in September 2005 at the University of Luxembourg. It is a continuation of the Bachelor studies as a first step towards the PhD. \textsf{MICS} currently starts with an orientation meeting where all new students get to know to the professors and other students. The first semester is then mandatory for all, because it concerns the fundamentals of computer science and acts as a preparation for the following semesters. By the end of the first semester, the student selects one specialization. The second and third semester offer specialized courses in the selected field, preparing the candidate for the final Master Thesis. 

\textsf{MICS} is completely taught in English and adheres to the Bologna agreement. \textsf{MICS} is taught by faculty members from the Computer Science and Communications Research Unit. Our primary mission is to conduct fundamental and applied research in the area of computer, communication and information sciences. Our goal is to push forward the scientific frontiers of these fields. Overall over 200 students pursue graduate and undergraduate degrees in Computer Science with individual supervision. Our students will acquire the necessary know-how for high-level research- and industry-oriented work. Currently we teach students from more than ten different countries, we have national and international cooperation agreements with universities across Europe, US, and China, and the private sector. The multilingual and intercultural environment empower students to work both individually and in multinational teams. For \textsf{MICS}, we are looking for ambitious students with a background in computer science or a relevant field, solid mathematical knowledge and a Bachelor or Diploma degree who are willing to study in a research-oriented field. To be eligible, a student should have a Bachelor of Science, a Bachelor of Engineering (with a significant part in computer science), or Diploma in Computer Science.

There will be no entry examination since the Core Courses act as orientation semester where the student has to demonstrate his/her competences. The eligibility of both Bachelor/Diploma degrees in other related fields and other degrees are to be decided individually by the \textsf{director of studies} of the Master program. The major advances are that \textsf{MICS} is a two-year full-time graduate Master’ program (120 ECTS). All courses are taught in English; there are no tuition fees, but an inscription fee of 100 Euro. Around 20h per week teaching is required, the courses take place at City of Luxembourg (Kirchberg).



\end{document}